\documentstyle[11pt,mrs2001,epsfig]{article}
\begin{document}
\title{The cosmic flow in the Local Supercluster: \\
\smallskip
Tracing PSCz tidal influence through optimized Least Action Principle}
  
\author{E. Romano-D\'{\i}az$^1$, E. Branchini$^2$ \& R. van de
  Weygaert$^1$.}  
\affil{$^1$Kapteyn Astronomical Institute, RuG, P.O. Box 800, 9700 AV
  Groningen, The Netherlands.}  
\affil{$^2$Universita' Degli Studi di Roma Tre, Via della Vasca Navale 84,
  00146  Rome, Italy.}

\begin{abstract}
  We assess the extent to which the flux-limited PSC$z$ redshift
  sample is capable of accounting for the major share of mass
  concentrations inducing the external tidal forces affecting the
  peculiar velocities within the Local Supercluster (LS). The
  investigation is based upon a comparison of the ``true'' velocities
  in 2 large N-body simulations and their reconstruction from
  ``observation-mimicking'' mock catalogues. The mildly nonlinear
  ``mock'' LS and PSCz velocities are analyzed by means of the Least
  Action Principle technique in its highly optimized implementation of
  Nusser \& Branchini's Fast Action Method (FAM). For both model
  N-body Universes, we conclude that the dipolar and quadrupolar force
  field implied by the PSCz galaxy distribution would indeed be
  sufficiently representing the full external tidal force
  field\footnote{a complete version of this work is reported in
    Romano-D\'{\i}az, E., Branchini, E., van de Weygaert, R., 2001,
    MNRAS., {\it submitted}}.
\end{abstract}
\begin{figure}[h]
\vskip -0.7truecm
\centering\mbox{\psfig{figure=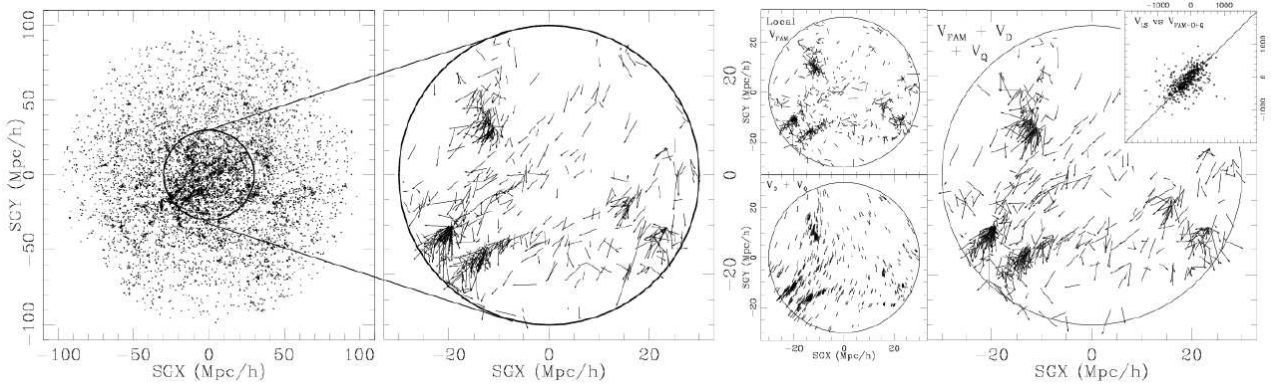,angle=-90,width=\hsize}}
\vskip -0.4truecm
\caption{Left: The analyzed ``mock'' Local Supercluster + PSCz galaxy 
  distribution, with the sample of ``mock'' measured Local
  Supercluster peculiar velocities. Right: FAM analysis of local
  velocity field: restricted to local cosmic volume (top left).
  Including full PSCz volume: external dipole and quadrupole
  components (D-Q, bottom left). Total of locally induced velocities +
  D-Q external contribution (right frame): agreement with full
  ``mock'' velocities illustrated by scatter plot (insert).}  
\vskip-1.0truecm
\end{figure}
\section{Cosmic flows: internal and external forces}
\vskip -0.25truecm
Migration flows of cosmic matter are one of the major physical
manifestations accompanying the emergence and growth of structure out
of the virtually homogeneous primordial Universe. Within the
gravitational instability scenario of structure formation, the
displacements are the result of the cumulative gravitational force
exerted by the continuously growing density surpluses and deficits
throughout the Universe.  This establishes a direct causal link
between gravitational force and the corresponding peculiar velocities.

Given a suitably accurate measurement of peculiar velocities within a
well-defined ``internal'' region of space, $V_{int}$, we may invert
these velocities, relate them to the gravitational force and whence
trace and possibly even reconstruct the source of the measured
motions. The gravitational force ${\bf g}({\bf x})$, \vskip
-0.25truecm
\begin{equation}
\quad {\bf g}({\bf x})\quad= \quad {\bf g}_{int}({\bf x})\ + \ {\bf g}_{ext}({\bf x})\quad = \quad {\bf g}_{int}({\bf x})\ +\ {\bf g}_D \ +\ {\bf Q}_{\bf g} \,\cdot\, {\bf x} \ + \ldots
\end{equation}
\vskip -0.15truecm
\noindent 
is the netto sum of the individual contributions from each cosmic location.   
We decompose the total integrated gravity ${\bf g}({\bf x})$ into an
internal component ${\bf g}_{int}$, representing the integrated
contribution from the density fluctuations within $V_{int}$, and the
externally induced gravitational force ${\bf g}_{ext}$, generated by
the fluctuations outside $V_{int}$. Often details of the external
matter distribution are not relevant for the dynamics in some local
region, so that sizable external force contributions are usually
restricted to the dipole and quadrupole components in ${\bf g}_{ext}$,
${\bf g}_D$ responsible for its ``bulk'' motion and ${\bf g}_{\bf
  Q}({\bf x})={\bf Q}_{\bf g} \,\cdot\, {\bf x}$, agent of the shear
in the flow field. In principle involving the complete visible
Universe, we expect significant contributions to be limited to some
finite volume whose typical scale will be that of the largest cosmic
structures. Although not exactly known, its size may amount to
$\approx 100-200h^{-1}\hbox{Mpc}$.  
\vspace{-0.5truecm}
\section{External influences in the Local Supercluster: a ``mock'' experiment}
\vspace{-0.35truecm}
We investigate the question of whether it is feasible to trace the
distant sources that have a noticeable influence on the dynamics in
our local Universe. Concretely, we focus our attention on the Local
Supercluster, and address the issue whether we can infer the extent
and configuration of the external gravitational forces influencing its
dynamics and evolution. In addition, we then to seek to assess whether
such a study provides sufficient information to identify the
responsible structures and objects from the present generation (or
shortly available) galaxy redshift samples.  To assess whether a
genuine observational study may yield conclusive evidence, we follow
the approach of studying idealized model circumstances.  Conclusions
are obtained from a comparison of the full model circumstances with
the predicted results from specially selected subsamples, moulded such
that they resemble the observational circumstances. For our model
Universes we take two large N-body simulations, one of a $\Lambda$CDM
structure formation scenario, the other of a $\tau$CDM scenario (Cole
etal. 1998).

Within each model, we {\it first} identify 10 ``mock'' Local
Superclusters, volume limited samples of galaxies in a spherical
volume of $30h^{-1}\hbox{Mpc}$ radius resembling the Nearby Galaxy
Catalogue (NBG). With these galaxy distributions at our disposal, we
determine the galaxy velocities that would result when gravity would
be solely the result of the internal mass distribution, i.e. assuming
a perfectly homogeneous external mass distribution. The deviations of
the predicted velocity field from the full velocities (here: the
N-body models) must then be ascribed to external forces.  In a {\it
  second} step we then try to evaluate where the most influential
external sources may be located. To this end, we use the PSCz galaxy
redshift survey as the template for our investigation of the
dynamically relevant external Universe. For the purposes of covering
uniformly all possible relevant matter concentrations, the 15,500
galaxies contained in this survey constitute an ideal sample. It is
based on an objectively defined IRAS flux selection, forms a near
perfect uniform flux-limited selection, has an excellent $84\%$ sky
coverage, and probes to a useful depth, $\sim 200h^{-1}\hbox{Mpc}$. On
the basis of its selection criteria we also identify, along with the
LS ``mock'' samples, 10 PSCz ``mock'' PSCz galaxy samples. The
lefthand frame of figure 1 depicts one specimen of our ``mock''
samples, showing the total LS+PSCz ``galaxy'' sample and zooming in
onto the ``LS region'', of which also the velocities are shown.

The Local Supercluster is a mildly nonlinear cosmic structure. Instead
of using simple linear dynamics or the Zel'dovich approximation we
therefore invoke the gravitational Least Action Principle (LAP,
Peebles 1989) for properly (re)tracing the galaxy orbits.  However,
the initial LAP implementation, suited for small galaxy groups, is not
adequate for our situation. The ``mock'' LS and PSCz galaxy samples
involves so many objects that the calculation would be rendered
impracticable. Possibly even more fundamental is its inability to deal
well with a flux-limited sample like the PSCz one. Of key importance
therefore is the use of the Fast Action Minimization (FAM)
optimization developed by Nusser \& Branchini (2000). Involving
numerical, force evaluation and base function optimizations at various
stages of the LAP technique, it allowed the large ``mock'' sample
computations in our analysis.

The results are highly encouraging. The righthand frame of figure 1
shows (topleft) the FAM velocity field prediction for the LS
restricted sample. Statistical analysis proved its incompatibility with
the full velocity field. But when the PSCz sample is invoked
significant differences vanish. Moreover, when restricting ourselves
to the corresponding dipole and quadrupole contributions, evaluations
on the basis of point-to-point velocity comparisons (see insert in
large righthand frame) prove these to represent virtually the complete
external contribution.
\vspace{-0.5truecm}
\section{Conclusions: ``what's in a game''}
\vspace{-0.35truecm} The FAM computations restrict themselves to the
matter concentrations within the realm of the PSCz selections and
prove that the implied velocities correspond extremely well with the
full ``model'' velocities. In other words, the PSCz selections appear
to comprise the source of the non-internally induced motions. Hence,
when we would live either in a $\Lambda$CDM or $\tau$CDM Universe, and
if the PSCz galaxies faithfully trace the underlying mass
distribution, we may plausibly argue the sample to contain all locally
relevant density concentrations. Moreover, almost all relevant
contributions can be ascribed to the dipolar and quadrupolar
anisotropy in the PSCz mass distribution. \vfill
\end{document}